\documentclass[preprint,12pt]{elsarticle}

\usepackage{epsfig}
\usepackage{makeidx}
\usepackage{graphicx}
\usepackage{epstopdf}

\usepackage{amssymb}

\usepackage[nodots]{numcompress}

\def\pb{\, .}
\def\vb{\, ,}

\newdefinition{rmk}{Remark}
\newproof{pf}{Proof}

\begin{document}

\begin{frontmatter}



\title{Micro to macro models for income distribution\\
         in the absence and in the presence of tax evasion}


\author{Maria Letizia Bertotti\corref{cor1}}
\ead{MariaLetizia.Bertotti@unibz.it}

\author{Giovanni Modanese}
\ead{Giovanni.Modanese@unibz.it}
 
\cortext[cor1]{Corresponding author}

\address{Faculty of Science and Technology \\
Free University of Bozen-Bolzano \\
Piazza Universit\`a 5, 39100 Bolzano, ITALY}

\begin{abstract}
We investigate the effect of tax evasion on the income distribution and the inequality index of a society through a kinetic model 
described by a set of nonlinear ordinary differential equations. The model allows to compute the global outcome 
of binary and multiple microscopic interactions between individuals. When evasion occurs, both individuals 
involved in a binary interaction take advantage of it, while the rest of the society is deprived of a part of the planned redistribution. 
In general, the effect of evasion on the income distribution is to decrease the population of the middle classes 
and increase that of the poor and rich classes. We study the dependence of the Gini index on several parameters 
(mainly taxation rates and evasion rates), also in the case when the evasion rate increases proportionally to a taxation rate 
which is perceived by citizens as unfair. Finally, we evaluate the relative probability of class advancement of individuals 
due to direct interactions and welfare provisions, and some typical temporal rates of convergence of the income distribution 
to its equilibrium state.
\end{abstract}


\end{frontmatter}



\section{Introduction}
\label{Intro}

The rise of inequalities in income and wealth, the implementation of different tax policies and the effects of tax evasion 
constitute important socioeconomic questions for most countries. Especially in times of economic crisis, like the current one, 
such issues become of major concern and are the object of frequent studies and debates. 
Involving a large number of interacting agents, as well as a multiplicity of aspects and levels, this matter certainly falls 
within the realm of the science of complex systems. We think that also mathematics can contribute to some extent 
to the analysis of these problems; for example, it can help to understand the micro-processes and mechanisms 
which lead to certain collective patterns. Through modelling and simulations, made possible by the power of modern computers, 
mathematics allows the exploration of several possible scenarios. Thus, in conjunction with the expertise from economics, 
political economics and other disciplines, and suitably supported by empirical data, mathematical models could 
in some cases even suggest concrete policies. 
 
Basically motivated by this belief,
we consider in this paper some microscopic models of taxation and redistribution 
in a closed market society, both in the absence and in the presence of tax evasion.
These models are constructed within a general framework 
which was first introduced in \cite{Bertotti M.L.} and then further investigated 
in \cite{Bertotti M.L. Modanese G. 1, Bertotti M.L. Modanese G. 2}.
They are formulated  as systems of nonlinear ordinary differential equations.
More precisely, the systems expressing them consist of a number $n$ of equations equal to
the number of income classes in which one divides a population.
The $j$-th equation (with $j = 1, ... , n$) describes the variation in time of the fraction, say $x_j$, of individuals belonging 
to the $j$-th class. The vector $(x_1, ... , x_n)$ represents the discrete income distribution over the population, whose size 
is supposed to remain constant. According to the findings of \cite{Bertotti M.L., Bertotti M.L. Modanese G. 1, Bertotti M.L. Modanese G. 2}, 
in correspondence to any value $\mu$ of the total income (which is a conserved quantity too)
a stationary income distribution exists, which is the asymptotic trend of all initial distributions with total income $\mu$.
In \cite{Bertotti M.L. Modanese G. 1, Bertotti M.L. Modanese G. 2} it was also shown that, if the number $n$ of classes is large enough 
($n$ was taken to be equal to $25$ in those papers) and if the value $\mu$ is compatible with initial distributions having 
the majority of individuals in the lower income classes, then for models constructed with suitable parameters 
(and those discussed here are of this type) the asymptotic income distributions exhibit fat tails 
with Pareto power-law behaviour like the real world distributions.

The main novelty with respect to the models explored in \cite{Bertotti M.L., Bertotti M.L. Modanese G. 1, Bertotti M.L. Modanese G. 2}
is that we treat here also cases in which tax evasion occurs. This addition
is not irrelevant.
Indeed, the illegal practice of tax evasion 
affects probably all societies, causing the ``loss'' of huge amounts of money,
which could be employed towards social and economic policies.
We are especially interested in the differences of the asymptotic 
income distributions in cases of tax compliance and in cases with tax evasion.
Below, we investigate these differences and we examine how
quantities and indicators like the Gini index, the tax revenue and the 
probability of class promotion due to welfare
change in the various cases.
In our approach the aggregate behaviour of a system, represented by the observable income distribution curves, 
emerges from the complex of interactions which take place between single heterogeneous individuals.

The underlying
behaviour- and interaction-based perspective differs intrinsically from the traditional viewpoint of 
mainstream economics, whose cornerstones are the assumption of a representative agent and the rational choice theory.
The interaction-based paradigm
began to take shape during the last decades and it counts among its pioneers 
various exponents of the economics community, e.g. T. Schelling, A. Kirman, B.W. Arthur, and M. Gallegati, 
see e.g. \cite{Arthur B. Durlauf S. Lane D.A., Gallegati M., Kirman A., Schelling T.C.}.
The tool kit of researchers adopting this perspective typically includes agent-based computational simulations
and complex networks. For example, questions related to tax evasion have been investigated via agent-based models 
in \cite{Bloomquist K.M., Hokamp S. Pickhardt M., Zaklan G. Westerhoff F. Stauffer D.}.
In these papers the focus is on the effect of interactions among behaviourally different agent types (honest, imitative, 
tax evaders and so on) on the changes in individual behaviour patterns. An experimental approach to such kind of questions
has been developed and described in \cite{Mittone L.}.

On another side, starting in the mid-1990s a branch of physics
denoted econophysics\footnote{\ The term {\it econophysics} was coined by H.E. Stanley.}
has been developed,
which explores the dynamical behaviour of economic and financial markets by means of methods and tools 
originally developed in statistical mechanics and in gas kinetic theory (see e.g. in this connection 
\cite{Chakraborti A. Chakrabarti B.K., Chatterjee A. Yarlagadda S. Chakrabarti B.K., 
Dragulescu A. Yakovenko V.M., Mantegna R.N. Stanley H.E., Sinha S. Chakrabarti B.K., Yakovenko V.M.}).
In econophysics the phenomenon of tax evasion has been described 
(e.g. in \cite{Zaklan G. Westerhoff F. Stauffer D.})
through an analogy with the Ising model, which is an array, typically 2-dimensional, of �spin� variables $s_{ij}$ that interact 
with their nearest neighbors and can only assume the values $\pm 1$. In the analogy each spin represents a citizen, 
which can be either in the �tax compliant� state $+ 1$ or in the �tax evader� state $- 1$ and can undergo transitions 
from $+ 1$ to $- 1$ due to imitation and from $- 1$ to $+ 1$ due to tax audits. 
Through numerical simulations or approximations typical of statistical mechanics it is possible to compute the average 
$\langle \sum_{i,j} s_{ij} \rangle$, directly related to the total evaders/compliant rate, as a function of several global or local parameters. 
This approach is helpful for the analysis of evasion phenomena in relation to local interaction and external controls, 
but not for studying the effect of evasion on the income distribution as we do here.

\smallskip

The paper is organized as follows. In the next section we sketch some
models, which were constructed and analysed in \cite{Bertotti M.L., Bertotti M.L. Modanese G. 1, Bertotti M.L. Modanese G. 2},
and we recall some of their features as established in these papers. In Section \ref{S incorporating tax evasion} we incorporate into these models
the tax evasion phenomenon and we discuss some first results concerning 
the asymptotic stationary income distributions 
which are found in the absence and in the presence of tax evasion.
We then explore in Section \ref{S proportional increase} the case in which to an increase of tax rates a proportional increase of evasion 
corresponds. The fifth 
and the sixth section are
devoted respectively to an in-depth analysis of some interesting quantities characteristic of the stationary solutions and
of the relative times of convergence.
Some summarizing comments are contained in Section \ref{S conclusion}.

\section{The tax compliance case}
\label{S tax compliance}

We shortly review in this section a family of models regarding a tax compliance case,
and we recall their main features as established in \cite{Bertotti M.L., Bertotti M.L. Modanese G. 1, Bertotti M.L. Modanese G. 2}.

\smallskip

Imagine dividing a population of individuals into a finite number $n$ of classes, each one characterized by its average income,
the average incomes being the positive numbers $r_1< r_2 < ... <\ r_n$. We refer to \cite{Bertotti M.L.} for a 
detailed illustration of the stylized micro scale mechanism we have in mind.
Here, we just recall that also the part of the government (which of course plays a role in connection with the taxation system)
can be described through monetary exchanges between pairs of individuals,
and we emphasise that consequently two kinds of interactions may take place: the so called {\it direct} ones, between an 
$h$-individual and a $k$-individual, occurring when the first one pays the second one,
and the {\it indirect} ones, between the $h$-individual and every $j$-individual
with $j \ne n$, occurring on the occasion of the direct $h$-$k$ interaction. The indirect interactions
represent the transactions corresponding to the payment of taxes and to the benefit of the redistribution.
In short, and we are referring here to a tax compliance case, in correspondence to any direct $h$-$k$ interaction, 
if $S$ (with $S < (r_{i+1} - r_{i})$ for all $i = 1, ..., n$)
denotes the amount of money that the $h$-individual should pay to the $k$-one, the overall effect of payment, 
taxation and redistribution is that of an $h$-individual paying a quantity $S \, (1 - \tau)$ to a $k$-individual 
and paying as well a quantity $S \, \tau$, which is divided among all $j$-individuals 
for $j \ne n$.\footnote{\ Individuals of the $n$-th class cannot receive money. Otherwise, they would possibly
advance to a higher class. And this is not permitted in the present context.} The quantity $\tau = \tau_k$, which is 
assumed to depend on the class to which the earning individual belongs, corresponds to the taxation rate of the $k$-th class.

A suitable framework towards modelling taxation and redistribution processes relative to such a population
was shown in \cite{Bertotti M.L.} to be provided by the system of $n$ nonlinear differential equations 
\begin{equation}
{{d x_i} \over {d t}} =  
\sum_{h=1}^n \sum_{k=1}^n {\Big (} C_{hk}^i + T_{[hk]}^i(x) {\Big )}
x_h x_k     -    x_i  \sum_{k=1}^n x_k \vb \qquad 
i= 1\vb ... \vb n \pb
\label{evolution eq eta = 1}
\end{equation}
Here, $x_i(t)$ with $x_i : {\bf R} \to [0,+\infty)$ denotes the fraction at time $t$
of individuals belonging to the $i$-th class; the coefficients $C_{hk}^i \in [0,+\infty)$, 
satisfying $\sum_{i=1}^n C_{hk}^i = 1$ for any fixed $h$ and $k$,
represent transition probability densities due to the direct interactions (more precisely, $C_{hk}^i $
expresses the probability density that an individual of the $h$-th class will belong to the 
$i$-th class after a direct interaction with an individual of the $k$-th class), and the functions
$T_{[hk]}^i : {\bf R}^n \to {\bf R}$, continuous and 
satisfying $\sum_{i=1}^n T_{[hk]}^i(x) = 0$ for any fixed $h$, $k$ and $x \in {\bf R}^n$, 
represent transition variation densities due to the direct interactions
(more precisely, $T_{[hk]}^i$ expresses the 
variation density in the $i$-th class
due to an interaction between an individual of the $h$-th class
with an individual of the $k$-th class).
The system $(\ref{evolution eq eta = 1})$
accounts for the fact that any direct or indirect interaction
possibly causes a slight increase or decrease of the income of individuals.

To choose a particular family of models, by specifying the values of the parameters $C_{hk}^i$ 
and the functions $T_{[hk]}^i(x)$, we first define certain coefficients $p_{h,k}$ for $h, k = 1, ... , n$.
These have the function of expressing the probability that  in an encounter between an $h$-individual and a $k$-individual,
the one who pays is the $h$-individual. Since also the possibility that none of the two pays has to be taken into account, the requirement 
which the $p_{h,k}$ must satisfy is that $0 \le p_{h,k} \le 1$ and $p_{h,k} + p_{k,h} \le 1$.
We take
$$
p_{h,k} = \min \{r_h,r_k\}/{4 r_n} \vb
$$
with the exception of the terms
$p_{j,j} = {r_j}/{2 r_n}$ for $j = 2, ..., n-1$,
$p_{h,1} = {r_1}/{2 r_n}$ for $h = 2, ..., n$, 
$p_{n,k} = {r_k}/{2 r_n}$ for $k = 1, ..., n-1$,
$p_{1,k} = 0$ for $k = 1, ..., n$
and
$p_{hn} = 0$ for $h = 1, ..., n$.
This choice, among others, was proposed and discussed in \cite{Bertotti M.L. Modanese G. 2}.

Our choice (see \cite{Bertotti M.L., Bertotti M.L. Modanese G. 2} for details) for the coefficients $C_{hk}^i$ 
and the functions $T_{[hk]}^i(x)$ is reported next. The only possibly nonzero elements among the $C_{hk}^i$ are:
\begin{eqnarray}
C_{i+1,k}^{i} & = 
                  & p_{i+1,k} \, \frac{S \, (1-\tau_k) }{r_{i+1} - r_{i}} \vb \nonumber \\
C_{i,k}^i & = 
            & 1 - \, p_{k,i} \, \frac{S \, (1-\tau_i)}{r_{i+1} - r_{i}} 
               - \, p_{i,k} \, \frac{S \, (1-\tau_k)}{r_{i} - r_{i-1}} \vb \nonumber \\
C_{i-1,k}^i & = 
               & p_{k,i-1} \, \frac{S \, (1-\tau_{i-1})}{r_{i} - 
	       r_{i-1}} \pb 
\label{choiceforC}
\end{eqnarray} 
We stress that the expression for $C_{i+1,k}^{i}$ in $(\ref{choiceforC})$ holds true for $i \le n-1$ and $k\le n-1$;
the second addendum of the expression for $C_{i,k}^i$ is effectively present only 
provided $i \le n-1$ and $k \ge 2$, while its third addendum is present
only provided $i \ge 2$ and $k \le n-1$; the expression for $C_{i-1,k}^i$ holds true for $i \ge 2$ and $k\ge 2$.

Following \cite{Bertotti M.L.}, we take the functions $T_{[hk]}^i(x)$ as
\begin{eqnarray}
T_{[hk]}^i(x) & = & \frac{p_{h,k} \, S \, \tau_k}{\sum_{j=1}^{n} x_{j}} {\bigg (}  \frac{x_{i-1}}{(r_i - r_{i-1})} -   \frac{x_{i}}{(r_{i+1} - r_{i})} {\bigg )} \nonumber \\
                    \ & + &  p_{h,k} \, S \, \tau_k \, 
{\bigg (} 
\frac{\delta_{h,i+1}}{r_h - r_{i}} \, - \, \frac{\delta_{h,i}}{r_h - r_{i-1}}
{\bigg )} 
\, \frac{\sum_{j=1}^{n-1} x_{j}}{\sum_{j=1}^{n} x_{j}} \vb 
\label{choiceforT}
\end{eqnarray} 
with $\delta_{h,k}$ denoting the {\it Kronecker delta}. In the r.h.s. of $(\ref{choiceforT})$, $h >1$ and the terms 
involving the index $i-1$ [respectively, $i+1$] are effectively present only provided $i-1 \ge 1$ [respectively, $i+1 \le n$].  

\rmk
We point out that,
due to the bound on the income of individuals in the $n$-th class,
in the model under consideration,
the effective amount of money representing taxes, which is paid 
in correspondence to a payment of $S (1 - \tau(k))$
and is then redistributed 
is $S \, \tau(k) \,({\sum_{j=1}^{n-1} x_{j}})/{(\sum_{j=1}^{n} x_{j}})$
instead of $S \, \tau(k)$.   

\medskip

To fix ideas, we take $S = 1$, 
\begin{equation}
r_j = 10 \, j \vb
\label{average incomes}
\end{equation}
and 
\begin{equation}
\tau_j = \tau_{min} +  \frac{j - 1}{n-1} \, (\tau_{max} - \tau_{min}) \vb
\label{progressivetaxrates}
\end{equation}
for $j = 1, ... , n$.
Still, the value of $\tau_{min}$ and $\tau_{max}$
have to be fixed. 
Hence, the equations $(\ref{evolution  eq eta = 1})$ describe a family of models rather than a single model.
They are well beyond analytical solutions. But, relevant facts can be understood through simulations. 
We notice that, to run simulations, we take $n=25$.

\rmk
It is worthwhile observing that the choices of the 
$C_{hk}^i$ and the $T_{[hk]}^i(x)$ in
$(\ref{choiceforC})$ and $(\ref{choiceforT})$ respectively are quite natural. In contrast, the choice of the $p_{h,k}$  
is an arbitrary one. We make it, because it seems to be reasonable and it guarantees some heterogeneity 
in the saving propensity
across classes.

\medskip

The following properties have been found to hold true in \cite{Bertotti M.L., Bertotti M.L. Modanese G. 1, Bertotti M.L. Modanese G. 2}.
Precisely, the first two have been analytically proved; the remaining ones are in fact suggested by a large number of simulations.

\medskip

$\bullet$ In correspondence to any initial condition $x_0 = (x_{01} , \ldots , x_{0n})$, 
for which $x_{0i} \ge 0$ for all $i = 1, ... , n$ and $\sum_{i=1}^n x_{0i} = 1$,
a unique solution $x(t) = (x_1(t),\ldots,x_n(t))$ of $(\ref{evolution eq eta = 1})$ exists,
which is defined for all $t \in [0,+\infty)$, satisfies $x(0) = x_0$ and also
\begin{equation}
x_{i}(t) \ge 0 \ \hbox{for} \ i = 1, ... , n \ \hbox{and} \ \sum_{i=1}^n x_{i}(t) = 1 \ \hbox{for all} \ t \ge 0 \, . 
\label{solution in the future}
\end{equation}
In particular, this implies that the expressions of the $T_{[hk]}^i(x)$ in (\ref{choiceforT}) simplify
and the right hand sides of system $(\ref{evolution eq eta = 1})$ are polynomials of degree $3$, see \cite{Bertotti M.L.}.

\smallskip

$\bullet$ The scalar function
$\mu(x)=\sum_{i=1}^n r_i x_i$, expressing the global (and mean) income, 
is a first integral for the system $(\ref{evolution eq eta = 1})$, see \cite{Bertotti M.L.}. 

\smallskip

$\bullet$ For any fixed value $\mu \in [r_1,r_n]$, 
an equilibrium of $(\ref{evolution eq eta = 1})$ exists, to which
all solutions of $(\ref{evolution eq eta = 1})$,
whose initial conditions $x_0 = (x_{01} , \ldots , x_{0n})$ satisfy
$x_{0i} \ge 0$ for all $i = 1, ... , n$, $\sum_{i=1}^n x_{0i} = 1$, and $\sum_{i=1}^n r_i x_{0i} = \mu$
tend asymptotically as $t \to +\infty$.
In other words, a one-parameter family of asymptotic stationary distributions exists,
the parameter being the total income value, \cite{Bertotti M.L., Bertotti M.L. Modanese G. 1, Bertotti M.L. Modanese G. 2}. 

\smallskip

$\bullet$ The profile of the asymptotic stationary distribution depends on
the difference between the maximum and the minimum tax rates, $\tau_{max} - \tau_{min}$. Specifically,
if this difference is enlarged, while all other data are kept unchanged,
an increase of the fraction of individuals belonging to the middle classes 
(to the detriment of those in the poorest and richest classes)
can be detected at the asymptotic equilibrium, \cite{Bertotti M.L., Bertotti M.L. Modanese G. 1}.

\smallskip

$\bullet$ The asymptotic stationary distributions corresponding to suitable values of the total income $\mu$
exhibit tails which have a power-law decreasing behaviour.
Such a property has been observed in real world income distributions since the work by
Pareto, \cite{Pareto V.}.
The condition on $\mu$ is related to the fact that the total income cannot be too high
if a tail is expected in the asymptotic distribution.
In practice, in the admissible distributions of the population,
the majority of individuals have to be concentrated in lower income classes.
Anyway, this occurs quite naturally in the real world, \cite{Bertotti M.L. Modanese G. 1, Bertotti M.L. Modanese G. 2}.

\section{Incorporating tax evasion into the model}
\label{S incorporating tax evasion}

In this section we incorporate in the model the occurrence of a partial tax evasion. 
Our first natural inspection is then devoted to comparing
the asymptotic income distributions which 
evolve from identical initial conditions
in the absence and in the presence of this illegal practice. 
Preliminar results in this direction have been
reported in \cite{Bertotti M.L. Modanese G. 3}.

The case we model is that one in which, the effect of tax evasion in a trade
corresponds to an advantage for both the individual who is receiving money and that one who is paying.
(Of course, other cases of tax evasion could be considered).
Such cases happen e.g. in connection with value added taxes.
Indeed, the payment of this kind of tax relies upon invoices and receipts.
A way how tax evasion may take place may be described as follows.
The individual who receives the money (in our scheme, a $k$-individual), 
be he an entrepreneur, a professional, a trader or similar,
colludes with the individual who is paying him (a $h$-individual),
offering a discount on condition that this will not require any invoice or receipt.
In this way, the $h$-individual has the advantage of the discount, and 
the $k$-individual
will conceal his gain from the tax return.\footnote{\ In such a case e.g. taxes called in Italy, our country, I.V.A. and I.R.P.E.F. are evaded.}

In order to keep into account, at least to some extent, such a behaviour, we recall that
according to Section \ref{S tax compliance}, in the case of tax compliance,
when an $h$-individual is supposed to pay an amount of money $S$ to a $k$-individual, 
what equivalently happens is that
\begin{itemize}
\item{the $h$-individual pays a quantity $S \, (1 - \tau_k)$
to the $k$-individual 

and he pays a quantity $S \, \tau_k$ to the government.}
\end{itemize}
We take now $\theta_k \le \tau_k$. To fix ideas, we take 
\begin{equation}
\theta_k = (1-q) \, \tau_k \vb \qquad \hbox{with} \ 0 \le q \le 1 \pb
\label{theta}
\end{equation}
We write the scaling coefficient in (\ref{theta}) as $1 - q$ so as to have that
the absence of evasion corresponds to $q = 0$ and
a total evasion corresponds to $q = 1$.
The effect of a partial tax evasion can be produced provided e.g. we postulate that,
when an $h$-individual is supposed to pay an amount of money $S$ to a $k$-individual, 
as a matter of fact
\begin{itemize}
\item{the $h$-individual pays a quantity $S \, (1 - (\tau_k + \theta_k)/2)$
to the $k$-individual 

and he pays a quantity $S \, \theta_k$ to the government.}
\end{itemize}
If $q > 0$, then the $h$-individual pays less than he should,
the $k$-individual gains in the end more than he would have done in the tax compliant situation
and the government collects less than it should.

\smallskip

Summarizing, in the presence of tax evasion the evolution equations are given by $(\ref{evolution eq eta = 1})$,
with the $C_{hk}^i$ and the $T_{[hk]}^i(x)$ as in $(\ref{choiceforC})$ and $(\ref{choiceforT})$, where however
$S \, (1 - \tau_k)$ is replaced by $S \, (1 - (\tau_k + \theta_k)/2)$ and
$S \, \tau_k$ is replaced by $S \, \theta_k$ 
for any $k = 1, ... , n$.

\smallskip

As described in \cite{Bertotti M.L. Modanese G. 3},
we ran several simulations, with different values of $q$,
relative to pairs of cases with the same initial conditions,
the first one without and the other one with tax evasion.

\begin{figure*}[h]
\begin{center}
\includegraphics[width=6cm,height=3cm] {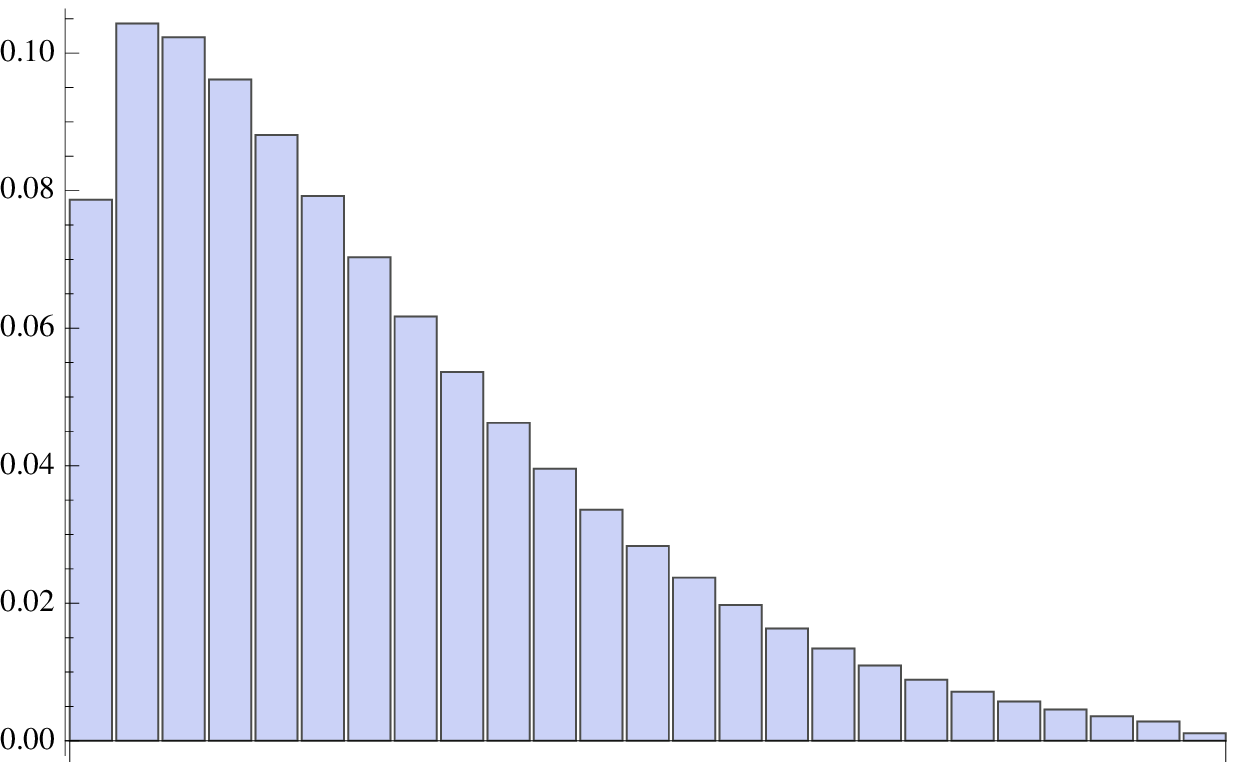}
 \hskip0.7cm
\includegraphics[width=6cm,height=3cm] {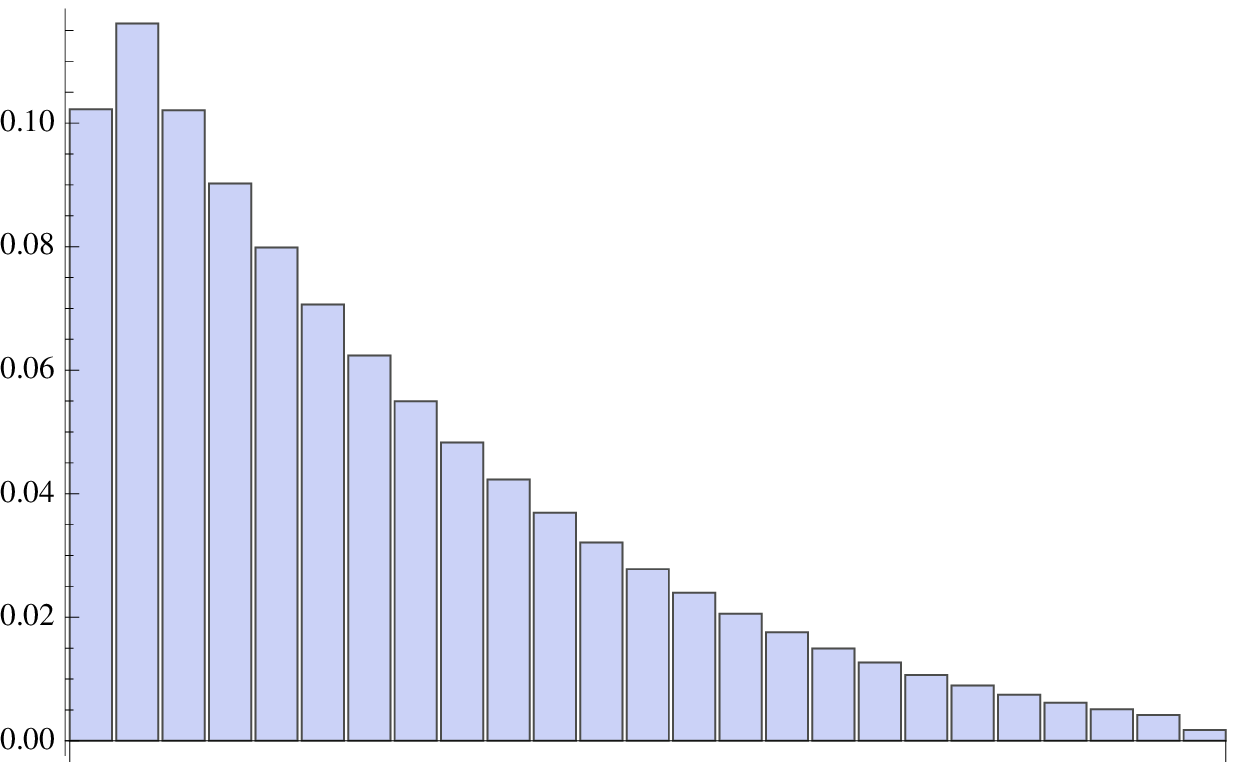}
\end{center}
\begin{center}
\includegraphics[width=6cm,height=3cm] {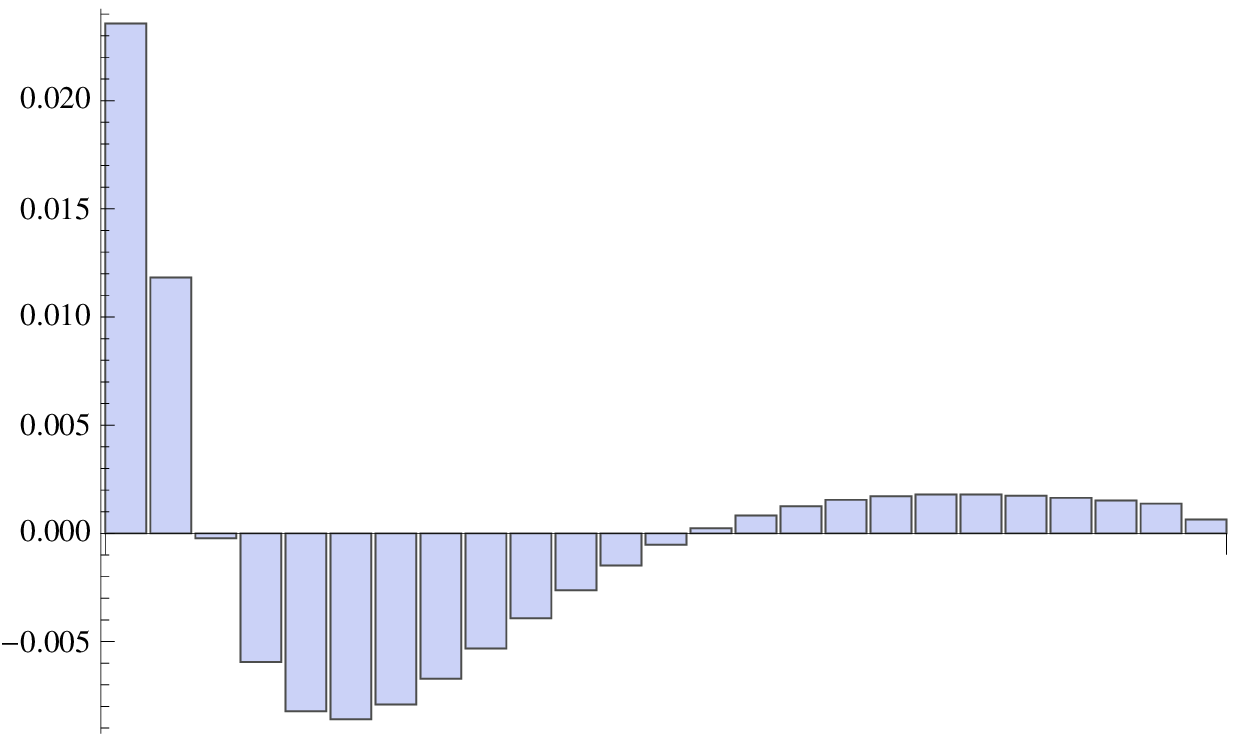}
 \hskip0.7cm
\includegraphics[width=6cm,height=3cm] {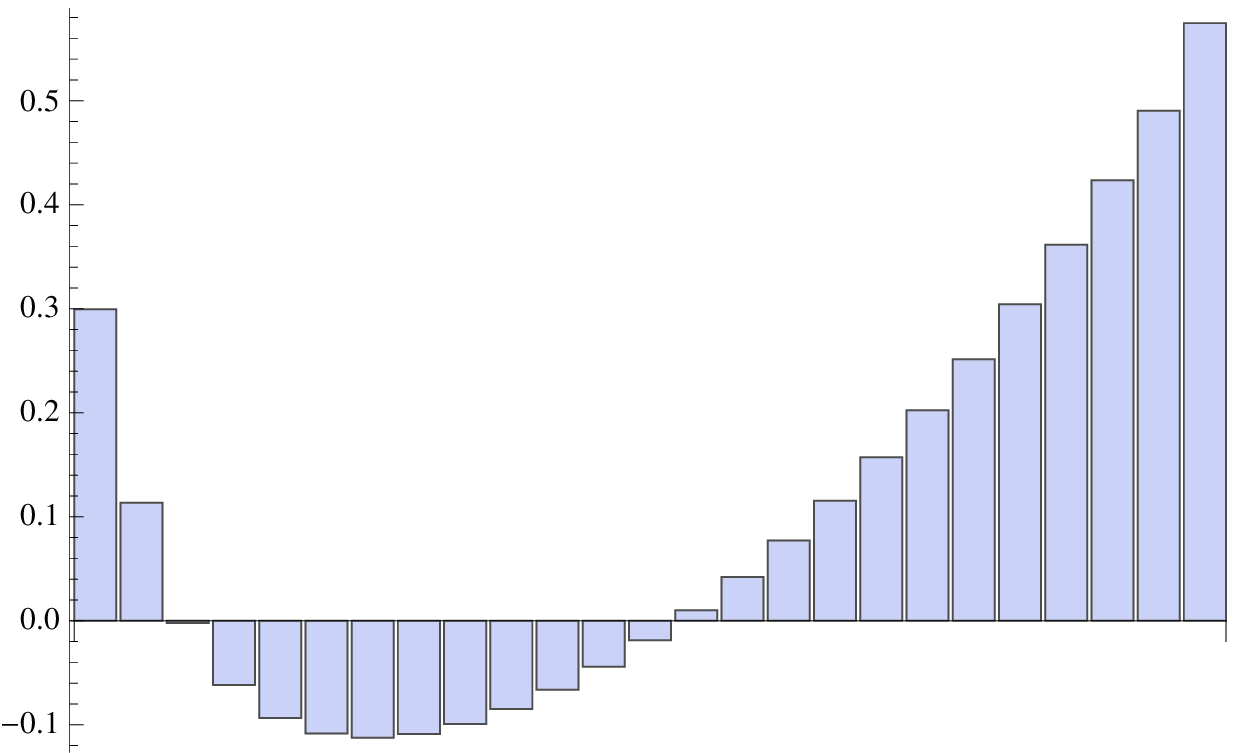}
\end{center}
\caption{In the first row, the panel on the left displays an asymptotic distribution in a case of tax compliance, while
the panel on the right displays the asymptotic distribution for the same initial condition in the presence of tax evasion
($\tau_{min} = 30 \%$, $\tau_{max}=45 \%$, $q = 1/3$);
in the second row,
the histograms in the panel on the left express the difference between the fraction of individuals in each class in the two cases with tax evasion 
and with tax compliance; the histograms in the right panel represent the percentual variation in each class
of the fraction of individuals when passing from the tax compliance case to the one with tax evasion.
The histograms are scaled differently on different pictures.}
\label{fig:confrontivirtuosoevasione}     
\end{figure*}

The simulations systematically show that the effect produced by tax evasion is 
an increment in the number of individuals belonging to the poorest and to the richest classes
at the detriment of the middle ones. Moreover, examining the percentage in each class
of the variation of the number of individuals, shows that, when passing from the straight to the dishonest case,
the increasing effect in the high income classes is larger for greater income,
while in the low income classes it is larger for lower income. Correspondingly, those who benefit from tax evasion
are individuals of the richest classes; and the richer they are, the most they benefit. In contrast, things are getting worse
for individuals with low income: many of them even pass to the lowest income class.
The four panels in the Figure \ref{fig:confrontivirtuosoevasione} illustrate the typical output.

Also the Gini index $G$, which provides a measure of the income inequality, turns out to be larger when tax evasion is present.
This index takes values in $[0,1]$, with $0$ representing complete equality and $1$ the maximal inequality. 
For the example in the Figure \ref{fig:confrontivirtuosoevasione}, it is approximately equal to $0.383$
in the tax compliance case and $0.410$ in the tax evasion case.
If the evasion rate is increased e.g. to $q = 2/3$, it is approximately equal to $0.444$, and so on.
We recall that the definition of $G$ involves the Lorenz curve, which plots the cumulative percentage of the total income of a population (on the $y$ axis) 
earned by the bottom percentage of individuals (on the $x$ axis).
Specifically, $G$ is the ratio $A_1/A_2$ of the area $A_1$ between the Lorenz curve and the line of perfect equality (the line at $45$ degrees)
and the total area $A_2$ under the line of perfect equality.
In our simulations we estimated the Gini index by calculating the area under the Lorenz curve as a sum of areas of trapezia.

\section{Proportional increase of tax rates and evasion}
\label{S proportional increase}

The extent of tax evasion in a society depends on several factors, like for instance the strength of moral values, imitation phenomena, 
frequency of the tax audits, strictness of the penalties etc. A further important factor which influences the decision of a citizen 
to attempt some form of evasion is his or her perception of the fairness and rationality of the taxation scheme, and the ``control'' that 
he or she can exert on the final destination and good use of the money collected through taxation, \cite{Mittone L.}.
In our model we can simulate quite easily a situation related to the perception of fairness, namely a situation in which the evasion rate 
increases in response to a sharp increase of the maximum tax rate $\tau_{max}$. 
Notice that, in view of $(\ref{progressivetaxrates})$, increasing $\tau_{max}$ produces an increase of all tax rates $\tau_k$ but $\tau_{min}$.
For simplicity we shall suppose that the evasion rate 
$q$ varies in the same way for all income classes, but it is straightforward to turn to the general case of class-specific evasion rates 
$q_k$.

As a first illustrative example, suppose to keep $\tau_{min}$ fixed, $\tau_{min}$ = 20 \%, while
varying $\tau_{max}$ and $q$ according to Table \ref{tab1}.
For each couple of values we compute the equilibrium distribution, its Gini index and also the 
tax revenue or government budget 
$W_{tot}$ (compare Section \ref{budget}). Our purpose is to see if there exists an ``optimal'' value of $\tau_{max}$ 
which allows, even in the presence of evasion, to minimize the inequality, while keeping at the same time the government budget in a reasonable range.

\begin{table}[h]
\begin{center}
\begin{tabular}{cccccc}
\hline\noalign{\smallskip}
$\alpha$ & & & $q \, (\%)$ & & $\tau_{max} \, (\%)$  \\
\noalign{\smallskip}\hline\noalign{\smallskip}
1\ & &&20 & & 40 \\
2\ & &&25 & & 45 \\
3\ & &&30 & & 50 \\
4\ & &&35 & & 55 \\
5\ & &&40 & & 60 \\
6\ & &&45 & & 65 \\
7\ & &&50 & & 70 \\
8\ & &&55 & & 75
 \\
\noalign{\smallskip}\hline
\end{tabular}
\end{center}
\caption{Example of a situation in which tax evasion increases in response to an increase of the maximum tax rate $\tau_{max}$. 
Note that $q=0$ corresponds to honest behaviour and $q=1$ to total evasion. Both $\tau_{max}$ and $q$ 
are expressend in percentual form. $\alpha$ is an integer index for reference to the graphs. The ratio $\Delta q / \Delta \tau_{max}$ is equal to $1$.}
\label{tab1}
\end{table}

From the point of view of real economics this representation may still look quite naive, but at the mathematical level it is far from trivial: 
we are dealing with a manifold of equilibrium solutions of 25 coupled non-linear equations which depend on two variable parameters 
(plus, of course, on the total income and on the other parameters which define the choice of a specific model within the entire ``family'').

\begin{figure*}[h]
  \begin{center}
  \includegraphics[width=6cm,height=3cm]  {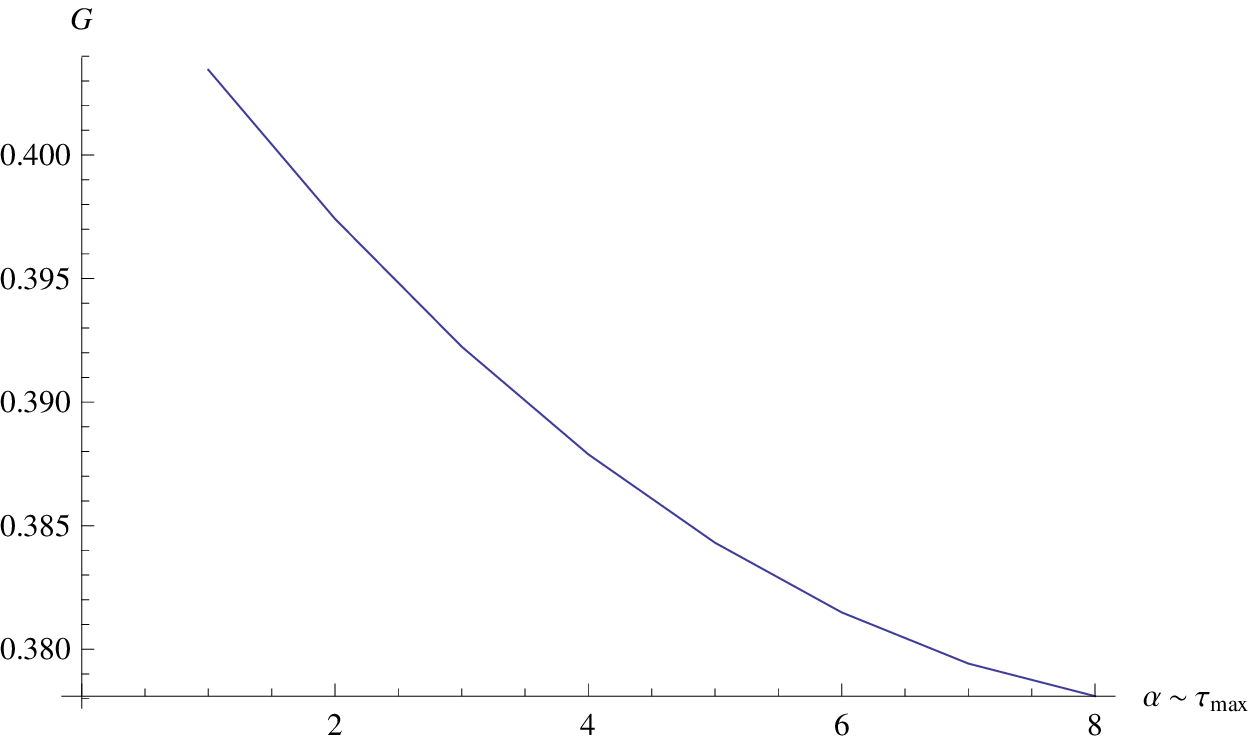}
  \hskip0.5cm
  \includegraphics[width=6cm,height=3cm]  {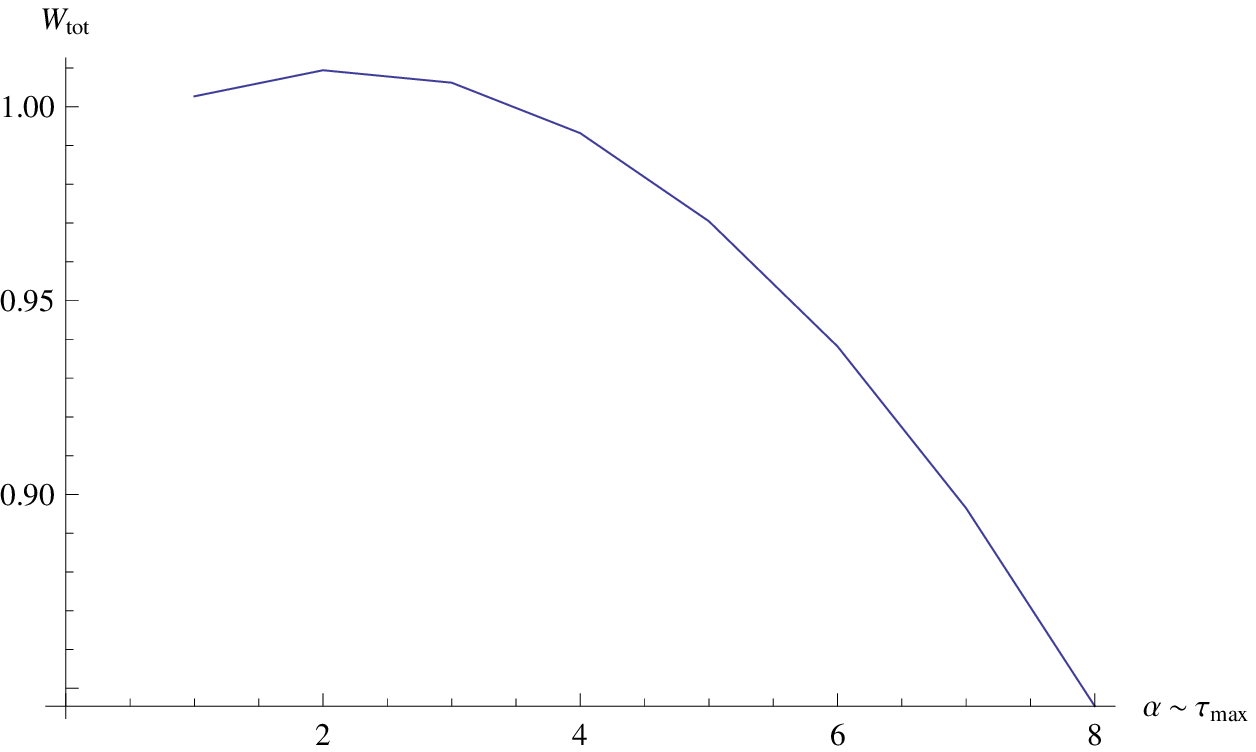}
  \end{center}
\caption{Behaviour of the Gini index $G$ and the government budget $W_{tot}$ as a function of $\tau_{max}$, 
for ratio $\Delta q / \Delta \tau_{max}$ equal to $1$ (compare Table \ref{tab1}). The function $G(\tau_{max})$ is decreasing. 
For graphical reasons the value of $W_{tot}$ has been multiplied by 100 here and in Fig. \ref{G-tau-2}.} 
\label{G-tau-1}       
\end{figure*}

Let us consider for simplicity only situations where $\tau_{max}$ and $q$ are varied keeping the ratio of their variation constant. 
In Table \ref{tab1}, for instance, the ratio is $1$. The resulting plot of the Gini index $G = G(\tau_{max})$ (Fig. \ref{G-tau-1}) is decreasing: this means that,
even though evasion increases with increasing taxation, the total effect is always a decrease of inequality (accompanied 
by sizeable variations in the total government budget). 

\begin{figure*}[h]
\begin{center}
\includegraphics[width=6cm,height=3cm]  {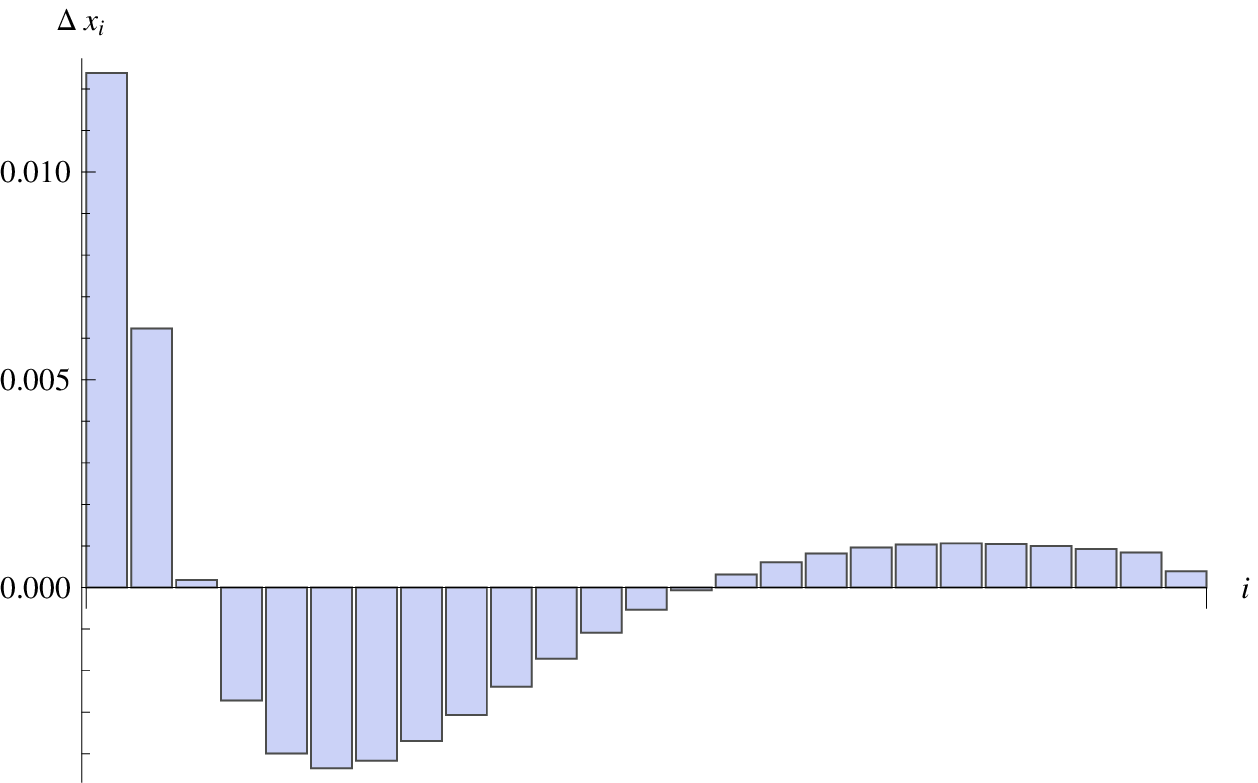}
\hskip0.5cm
\includegraphics[width=6cm,height=3cm]  {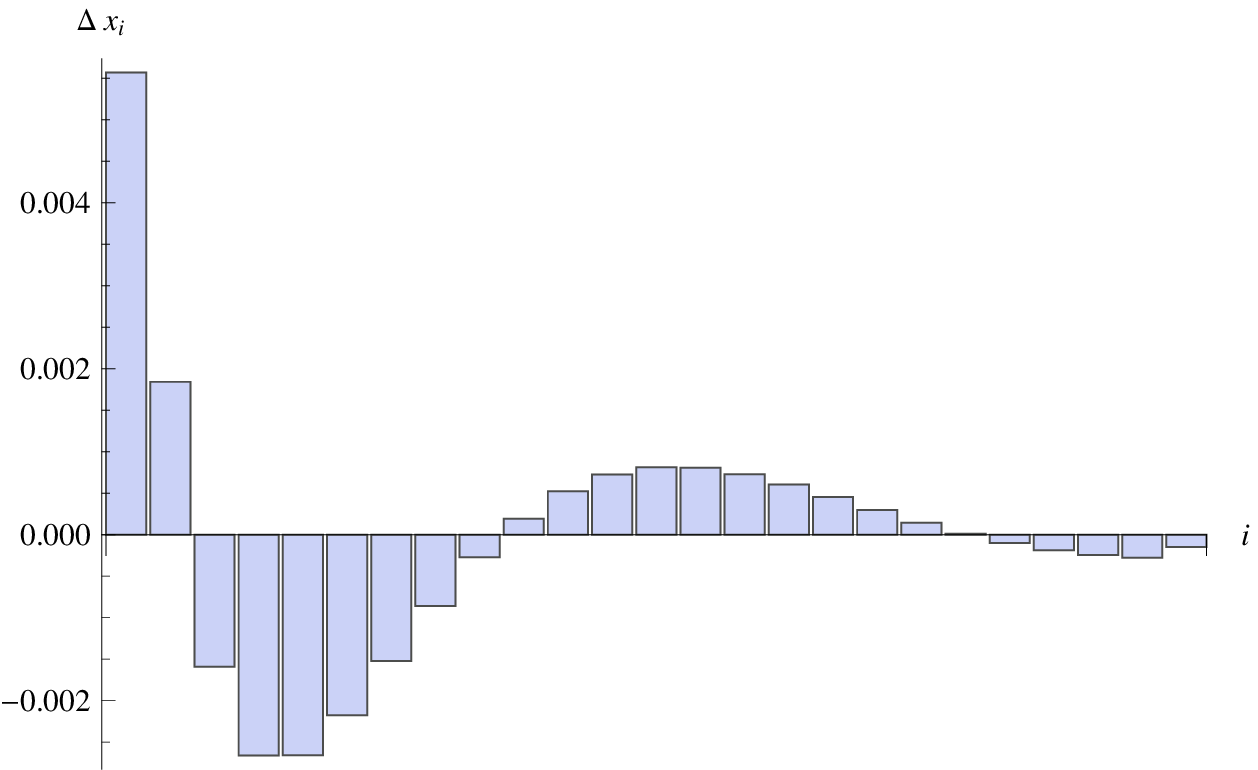}
\end{center}
\caption{Typical population variation for each income class due to the occurrence of evasion, in cases of variable $q$ 
and fixed $\tau_{max}$ (left panel) and in cases of fixed $\Delta q / \Delta \tau_{max}$ (right panel). 
In the case of the right panel, the middle class is split in two, and the super-rich decrease.} 
\label{onde}       
\end{figure*}

If we look in detail at the population variations for each income class with respect to the case without evasion, we note a curious phenomenon, 
which did not occur in the solutions with constant $\tau_{max}$. When $\tau_{max}$ grows, at first the number of the poor and rich increases, 
while the middle class shrinks, as it happens when there is evasion for fixed $\tau_{max}$ (Fig. \ref{onde}, left panel); at some point, however, 
for certain values of the ratio $\Delta q / \Delta \tau_{max}$,
the super-rich begin to decrease and the middle class is split in two sections with different trends: 
a middle-rich section, increasing, and a middle-poor section, decreasing (Fig. \ref{onde}, right panel). 

\begin{figure*}
\begin{center}
\includegraphics[width=6cm,height=3cm]  {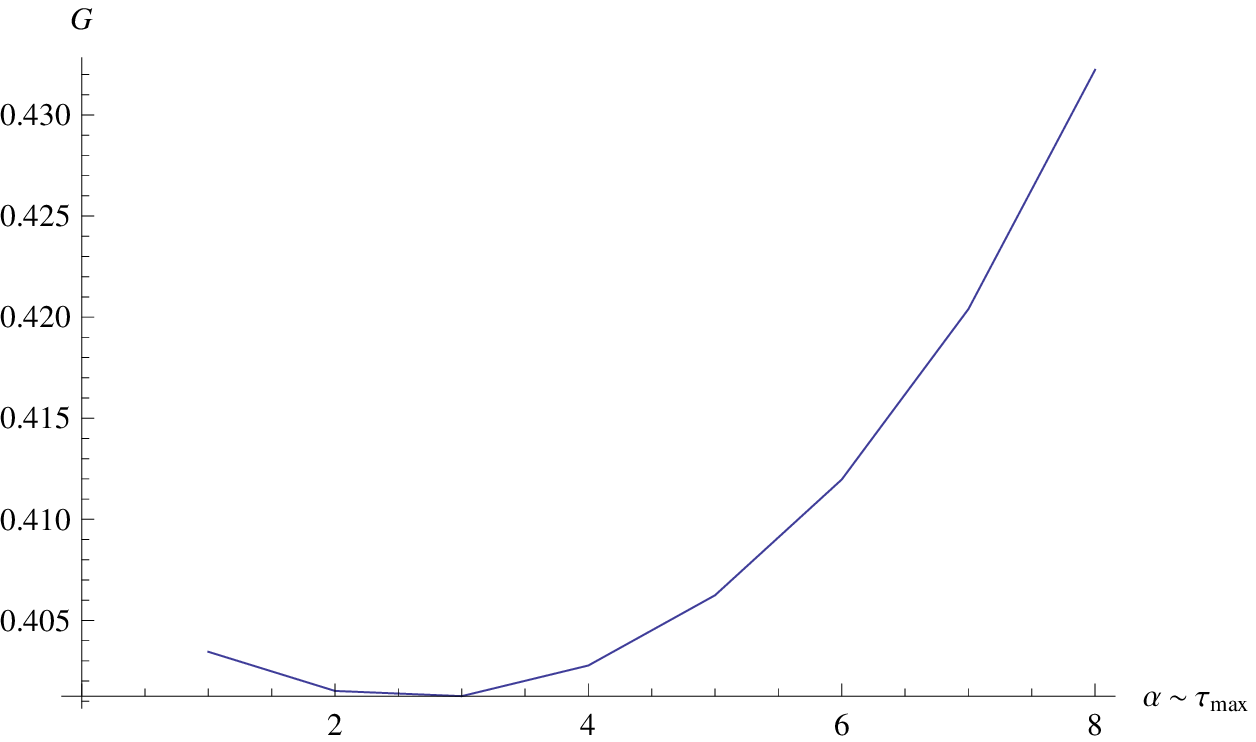}
\hskip0.5cm
\includegraphics[width=6cm,height=3cm]  {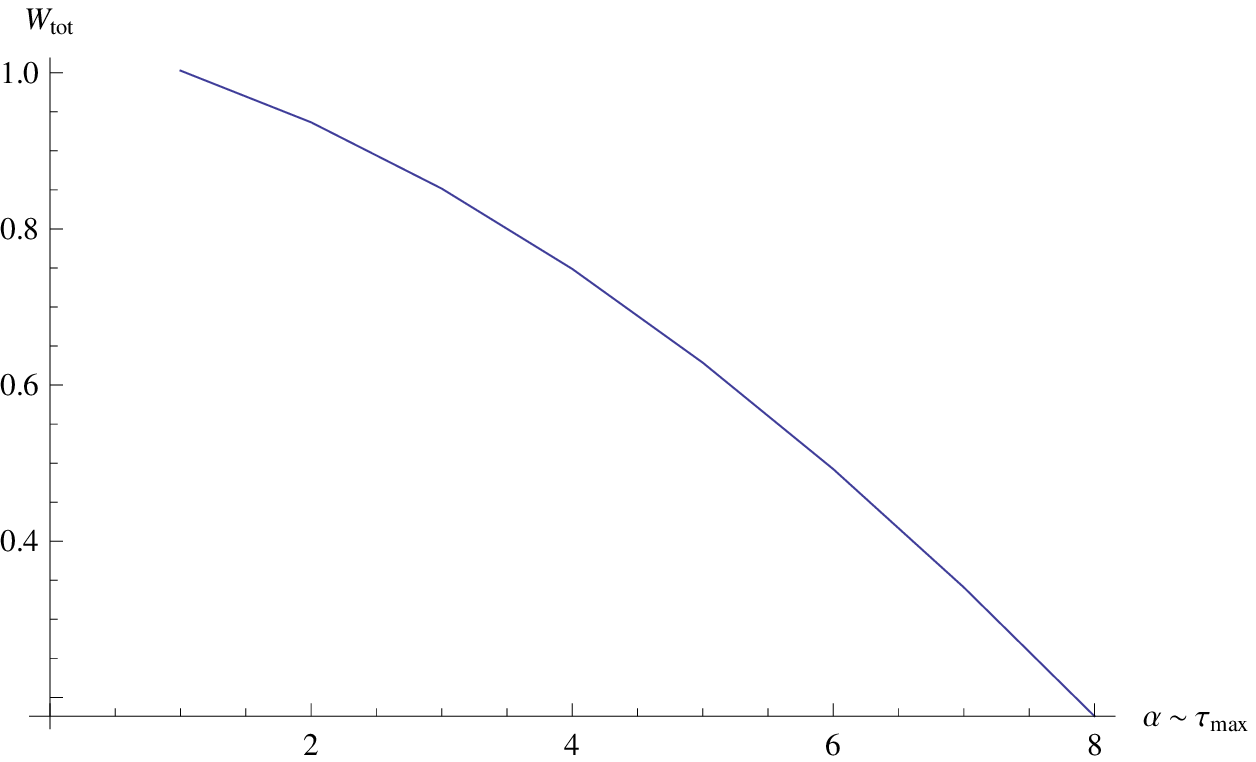}
\end{center}
\caption{Behaviour of the Gini index $G$ and the government budget $W_{tot}$ as a function of $\tau_{max}$, 
for ratio $\Delta q / \Delta \tau_{max}$ equal to $2$. The function $G(\tau_{max})$ has a minimum. This corresponds to 
an ``optimal'' value of $\tau_{max}$ which allows, even in the presence of evasion, to minimize the inequality 
while keeping at the same time the government budget in a reasonable range.} 
\label{G-tau-2}       
\end{figure*}

If, however, we choose a markedly different value for the ratio $\Delta q / \Delta \tau_{max}$, the behaviour of the Gini index is altered quite radically, 
and a minimum appears in the graph of $G(\tau_{max})$ (Fig. \ref{G-tau-2}). The minimum is clearly visible, for instance, 
when $\Delta q / \Delta \tau_{max} = 2$, i.e. when the evasion grows twice as fast in response to higher tax rates, compared 
to the case of Table \ref{tab1}. It is natural to ask for which value of the $\Delta q / \Delta \tau_{max}$ ratio the minimum begins to appear. 
The numerical solutions indicate that this happens for $\Delta q / \Delta \tau_{max} \simeq 1.1$. Of course, this value does not have 
an absolute meaning, since it still depends on the $p_{h,k}$ parameters of the model, on $\tau_{min}$ and on the initial value 
arbitrarily chosen for $q$ in the variations (here, $q = 0,2$). Yet the existence of ``phases'' with different behaviour in the parameter space is quite clear.

\section{Further ``dynamical'' quantities characteristic of the equilibrium}
\label{S further}

The equilibrium income distribution resulting from our model in correspondence to certain given parameters is completely described by the histogram 
of the equilibrium class populations $\left\{ \hat x_1,...,\hat x_{25} \right\}$. This histogram can be compared to real statistical data or to suitable fit functions, 
like the Gibbs function, the log-normal function or the Kaniadakis function (see \cite{Kaniadakis G.} and  \cite{Bertotti M.L. Modanese G. 2}
and references therein),
also in order to check for the possible presence of ``fat'' power-law tails. 

Some integral indices of the income distribution are usually computed in dependence of the model parameters and allow a quick comparison to real data. 
We have analyzed in Section \ref{S incorporating tax evasion} for instance, the dependence of the Gini inequality index $G$ 
on the variations of the evasion parameter $q$ 
and then in Section \ref{S proportional increase} the dependence of $G$ of simultaneous variations of $\tau_{max}$ and $q$. 
Further integral indices which are usually computed for the distribution functions of statistical physics are 
the average $\langle x \rangle$ and the variance $\sigma_x^2$. In our case, however, the average income is not meaningful 
because it is fixed by the initial conditions, and the variance does not appear to be of special interest. 

There are, nevertheless, some other interesting and peculiar integral quantities which characterize the equilibrium state 
of our model and depend on the choice of its parameters. Such quantities cannot be computed only from the 
asymptotic stationary distribution function, 
but depend on the underlying dynamical structure of the model. In order to define them, let us first recall that the equilibrium state is, 
like in any kinetic model, the result of a dynamical equilibrium: the number of individuals belonging to a certain class 
remains constant in time when the total rate of individuals leaving that class is equal in absolute value to the total rate 
of individuals arriving from other classes. Each single rate contains in turn contributions due to direct interaction and to taxation and redistribution. 

\subsection{The tax revenue}
\label{budget}

The simplest dynamical integral quantity is the total amount of tax collected in the unit time and redistributed as welfare provisions. 
It is often called tax revenue or government budget and it is given by
\begin{equation}
W_{tot} = \sum_{h=1}^{n} \, \sum_{k=1}^{n} \,\sum_{j=1}^{n - 1} \, S \, p_{hk} \, \theta_k \, {{\hat x}_j}{{\hat x}_h}{{\hat x}_k} \vb
\end{equation}
where ${\hat x}_i$ is the class population at equilibrium. 
On one hand, it is then interesting
to compare this budget with the total amount of the private-sector direct exchanges (see Section \ref{relative}).
On the other hand, it is also important, 
to keep track of the variations of the government budget when different taxations schemes are simulated. 
We have seen, for instance, that by increasing the gap $\tau_{max} - \tau_{min}$ between the maximum and minimum tax rates 
one typically obtains an income distribution where the middle classes are more populated and the Gini inequality index is smaller. 
This does not automatically imply, however, that the total amount of the collected taxes also increases; $W_{tot}$ might as well 
stay constant or decrease, and this would have significant consequences for the public sector.

Another typical case where one should be careful to keep the government budget under control while simulating changes of taxation is 
that of an increase of evasion following the increase of the maximum tax rate (Section \ref{S proportional increase}). Supposing that 
some amount of evasion is inevitable in practice, we have been looking for values of the 
evasion-taxation variation ratio $\Delta q / \Delta \tau_{max}$ which yield a minimum $G$ index. In correspondence to such values, 
one can in principle obtain a situation of relatively low inequality even in the presence of evasion. But what about the government budget? 
Will it still be sufficient to keep the state administration and welfare working, without the need for major cuts? If this is not the case, 
then one should conclude that social equality and evasion are incompatible and that the government must in any case enforce 
tax compliance by introducing further audits, fines etc. Consider for instance the case of the Figure \ref{G-tau-2}.

\subsection{Relative probability of class promotion due to welfare or direct interaction}
\label{relative}

The probability of class promotion per unit time following the interaction of an individual with others is given by the ratio 
between the money gained in the interaction and the income difference of the classes. This probability has therefore various contributions; 
some represent the money gained in direct binary interactions and others the money gained in indirect interactions due to taxation 
and welfare redistribution, represented in our model by terms of degree 3 in the population densities $x_i$. It is interesting to compute 
the ratio of these two contributions. Suppose, for instance, that for a certain class the total promotion probability per unit time at equilibrium 
is 0.15, of which 0.1 is due to direct interactions and 0.05 to indirect interactions. We can conclude that each individual of that class 
receives on the average, in the unit time, a certain amount of money from direct economic interactions, and half as much 
in the form of government welfare. If with different model parameters the ratio was, say, 1/10 instead of 1/2, then we could conclude 
that with those parameters the model represents a more ``liberistic'' society, and so on. We can also sum the probabilities over all classes, 
before computing that ratio. In that way we obtain a figure referred to the entire society, namely the ratio between the total amount collected 
and redistributed by the government through welfare schemes and the total amount of direct exchanges between individuals.

In conditions of dynamical equilibrium, the probability 
$P_{i,welfare}$
of class promotion 
due to welfare of an individual of the class $i$, for $1 \le i \le n - 1$, is
\begin{equation}
{P_{i,welfare}} = \frac{S}{{{r_{i + 1}} - {r_i}}} \, \sum_{h=1}^{n} \, \sum_{k=1}^{n} \, {{p_{h,k}}{\theta _k}{{\hat x}_h}{{\hat x}_k}} \vb
\end{equation}
where ${\hat x}_i$ is the class population at equilibrium. 
The probability $P_{i,exchanges}$ of class promotion due to direct exchanges is obtained from the appropriate 
terms of the matrix $C^i_{hk}$, i.e. is
\begin{equation}
{P_{i,exchanges}} = \frac{S}{{{r_{i + 1}} - {r_i}}} \,\sum_{k=1}^{n} \, {{p_{k,i}}\left( 1-\frac{\theta _i+\tau_i}{2} \right){{\hat x}_k}} \pb
\end{equation}

\begin{figure*}
\begin{center}
\includegraphics[width=6cm,height=3cm]  {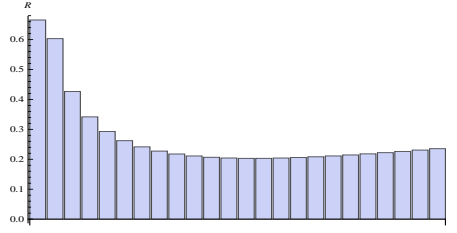}
\end{center}
\caption{An example 
of the ratio ${\it R} = {P_{i,welfare}}/{P_{i,exchanges}}$, giving the relative probability for 
an individual of the class $i$ to be promoted to the upper class due to welfare provisions or direct exchanges. 
(Here $\tau_{min}=20$, $\tau_{max}=50$, $q=0.3$; the coefficients $p_{h,k}$ are proportional to $\min (r_i,r_k)$).} 
\label{ratioPP}       
\end{figure*}

The mentioned ratio ${P_{i,welfare}}/{P_{i,exchanges}}$, computed in a special case, is represented in the Figure \ref{ratioPP}. 
From this graph we may deduce that for the very poor the welfare is an important factor of social promotion, while in the middle classes 
it becomes less important, in comparison to direct exchanges. What is quite surprising, is that the ratio stays almost constant 
when we pass to the rich classes, although one might expect welfare provisions to be quite irrelevant for the class advancement 
of the super-rich. A possible explanation is that in our kinetic model the direct exchanges involving the super-rich are quite rare, 
while all the tax money is constantly redistributed to everybody except to individuals of the $n$-th class (not represented in Figure \ref{ratioPP}). 
A more realistic version might take into account the fact that 
welfare benefits are usually not accessible to the super-rich.

\section{The time evolution scale}
\label{S time evolution scale}

\begin{figure}
  \begin{center}
\includegraphics[width=8cm, height=5cm]{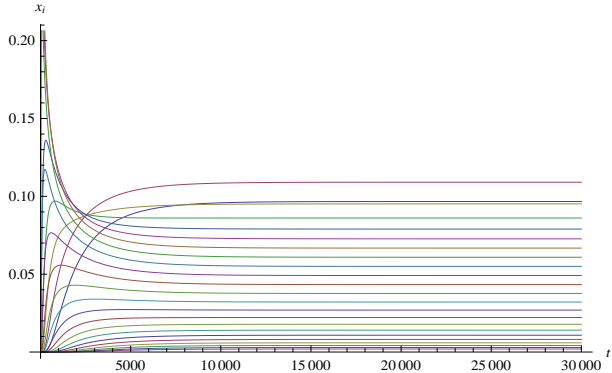}
\caption{Time evolution of the class populations $x_i$ from ``extreme'' initial conditions where all individuals 
are in the same income class, the class nr. 7. (Total income $\mu=70$; $\tau_{min}=30$, $\tau_{max}=50$, $q=0.5$, 
coefficients $p_{h,k}$ proportional to $\min (r_i,r_k)$.)}
  \end{center}
\label{lenta}       
\end{figure}

In our model the equilibrium income distribution is obtained from the numerical solutions of the differential equations at large times. 
The convergence of the solution to its equilibrium value is apparent from the numerical values of the $x_i(t)$ and from their temporal graphs. 
In general, the various class populations converge to their equilibrium values in different times, depending on ``how far'' 
they were from those values at the beginning. This is clear from the simultaneous graph in time of all the $x_i$ components 
(Fig. \ref{lenta}),
where some of their lines can be seen crossing each other at different instants.

\begin{figure}
  \begin{center}
\includegraphics[width=8cm, height=5cm]{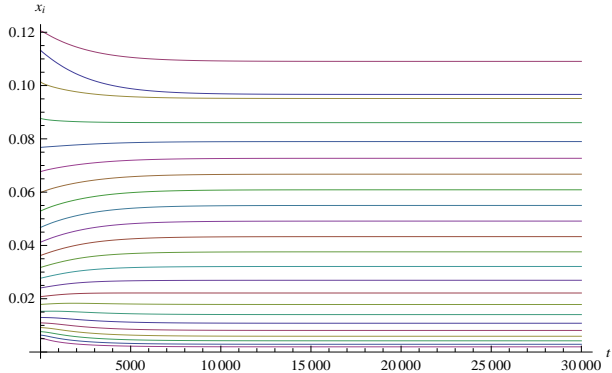}
\caption{Example of time evolution for a society already at equilibrium which adapts to a moderate change in the taxation rate. 
We have taken the equilibrium state of Fig. \ref{lenta} as new initial condition and changed $\tau_{min}$ to 10 and $\tau_{min}$ to 70, 
all the other parameters being the same.}
  \end{center}
\label{veloce}       
\end{figure}

\begin{figure}
\begin{center}
\includegraphics[width=6cm,height=3cm]  {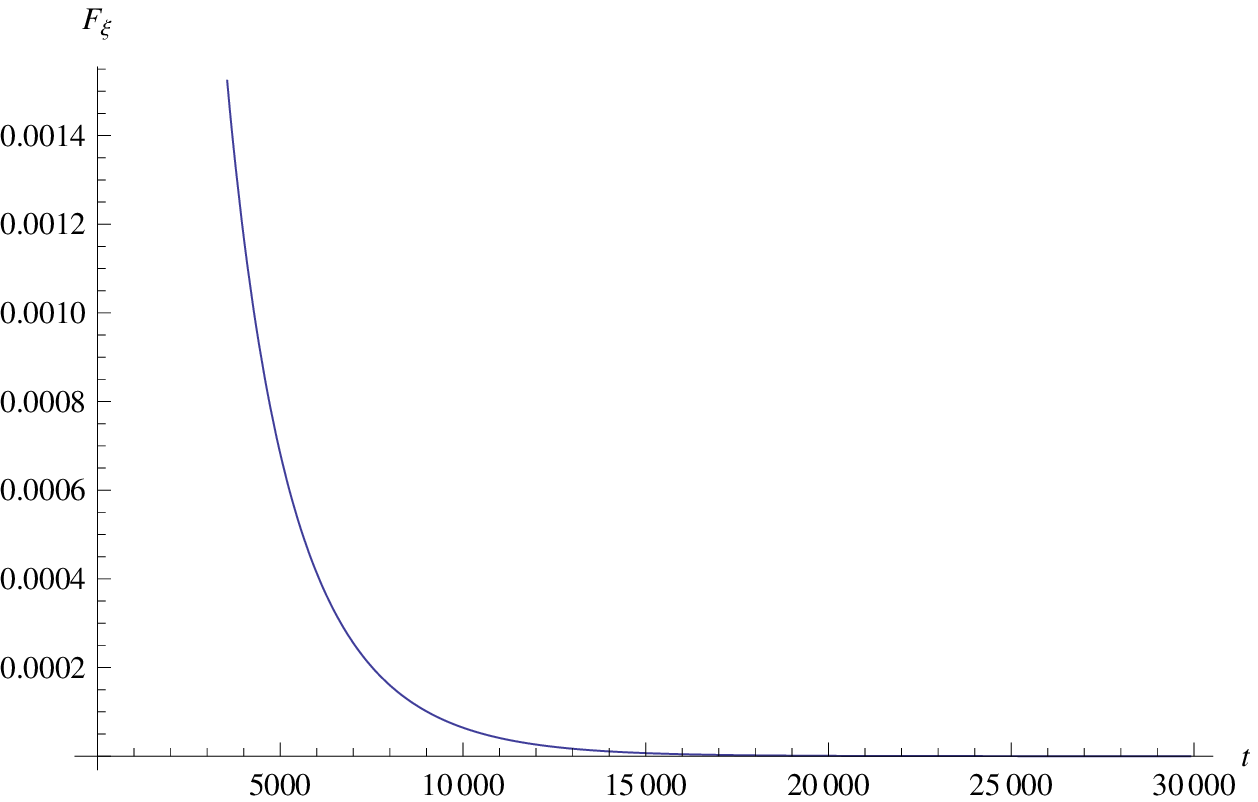}
\hskip0.5cm
\includegraphics[width=6cm,height=3cm]  {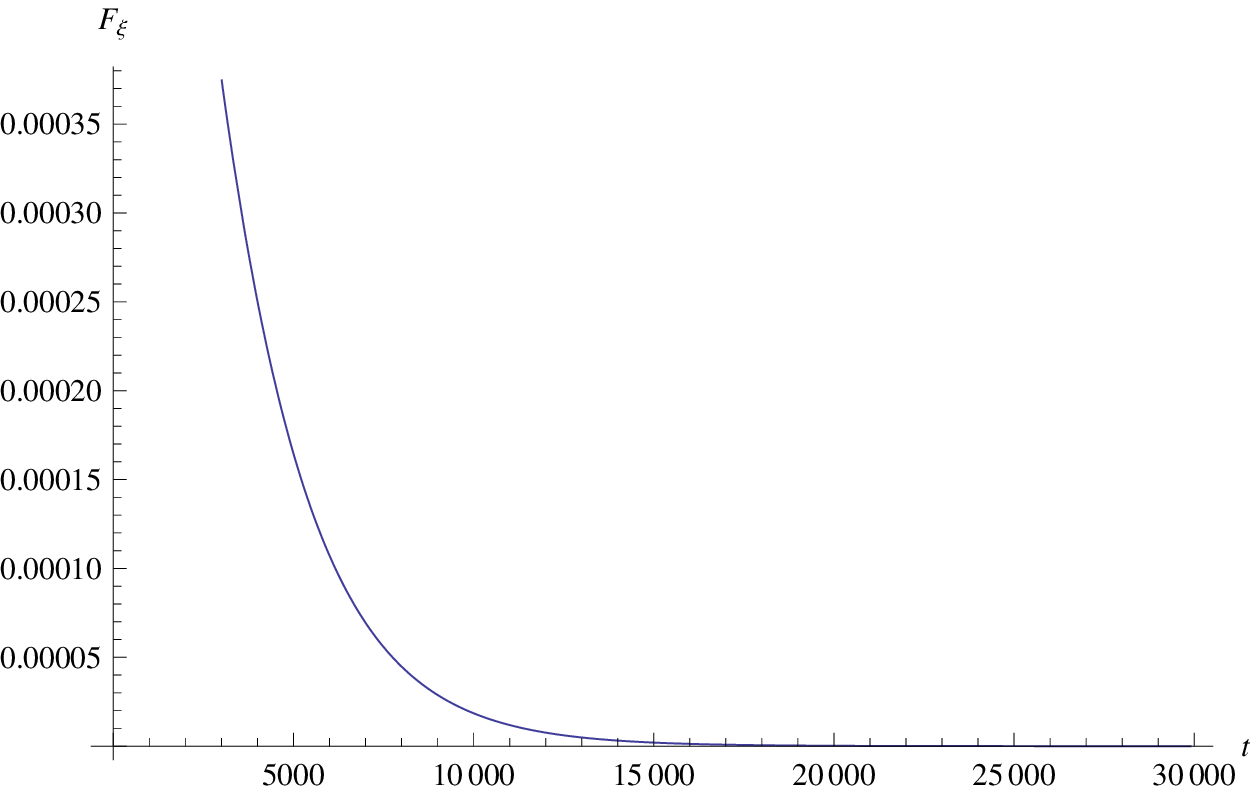}
\end{center}
\caption{Convergence norm $F_\xi(t)$ computed with $\xi=100$ for the two examples of slow evolution (left panel) and fast evolution (right panel) shown respectively 
in Figs. \ref{lenta}, \ref{veloce}. At the same time, the norm in the right panel is approximately ten times smaller than in the left panel. Due to the exponential behaviour, 
if we fix a threshold $\varepsilon$, the time $T$ necessary to reach it is approximately 
twice as large for the case in the left panel, compared to that in the right panel; for instance, 
for $\varepsilon=10^{-4}$ one has $T_a \simeq 11000$, $T_b \simeq 5500$.}
\label{confronto-norme}       
\end{figure}

\begin{figure}
\begin{center}
\includegraphics[width=6cm,height=3cm]{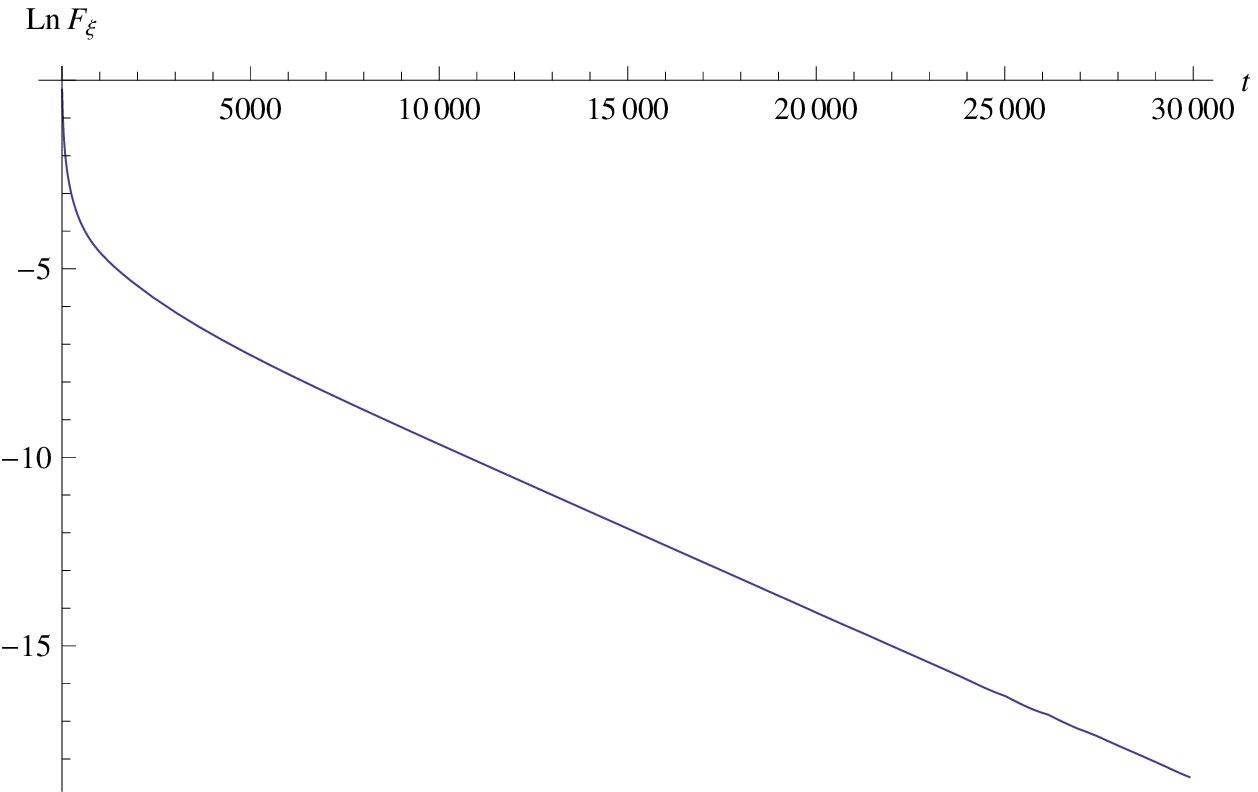}
\hskip0.5cm
\includegraphics[width=6cm,height=3cm]{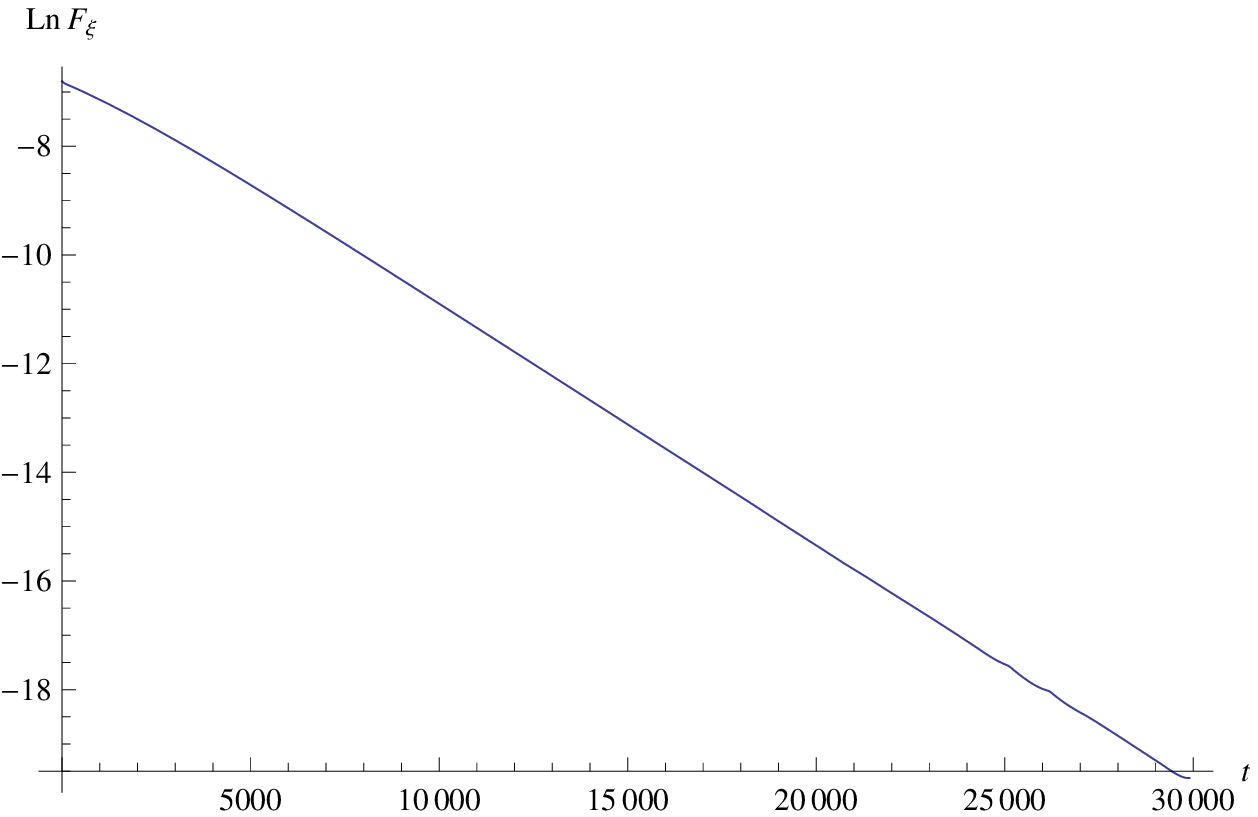}
\end{center}
\caption{Logarithm of the norm $F_\xi(t)$ in the case of slow and fast evolution in Fig. \ref{confronto-norme}, left and right panel. 
The exponential convergence of the norm is apparent.}
\label{log-norma}       
\end{figure}

In order to obtain an estimate of the temporal convergence which is uniform with respect to the various income classes, 
one can consider a vector norm applied to the difference between the configuration at time $t$ and the configuration at time $t+\xi$, being $\xi$ an arbitrary fixed delay;
i.e. one can consider the function
\begin{equation}
F_\xi(t) = \left[ \sum_{i=1}^{25} \left( x_i(t)-x_i(t+\xi) \right)^2 \right]^{1/2} \pb
\end{equation}
This reminds of the Cauchy convergence criterum for discrete successions.
The plot of $ F_\xi(t)$ converges quickly to zero as $t \to \infty$ (compare Fig. \ref{confronto-norme}). The plot of $\ln \, (F_\xi(t))$ in function of $t$ is linear 
(Fig. \ref{log-norma}), thus showing that the convergence of $ F_\xi(t)$ to zero is exponential. In correspondence of any small $\varepsilon$ 
one can find a convergence time $T$ such that $F_\xi(t) \leq \varepsilon$. This time also depends on $\xi$ and does not have 
any special significance in itself, but it allows to compare situations with slow and quick convergence. For instance, 
one of the 
longest convergence times, which we could consider as a reference time for our system, is the one obtained 
with the ``artificial'' initial conditions where all individuals are in the same income class (Fig. \ref{lenta}).
If we could actually ``reset'' a real society in this way and record the subsequent evolution, the convergence time would be likely of the order of years. 
This can be compared with the time it takes for a society already at equilibrium to adapt to a moderate change in the taxation rate (Fig. \ref{veloce}).

\section{Conclusion}
\label{S conclusion}

In this paper a microscopic model for the complex of monetary exchanges, taxation and redistribution in a closed society is investigated,
both in the tax compliance case and in the presence of tax evasion. 
The focus
is on the effects of
the tax evasion phenomenon 
on the income distribution over the population. Various comparisons 
between the situations without and with tax evasion
are made
through a direct inspection of the asymptotic income profiles,
and
by means of indicators like the Gini index, the tax revenue and the welfare-induced probability of class promotion.
In a nutshell, this stylised model supports the belief that a 
fair fiscal policy and a honest behaviour of the population individuals 
play a decisive role towards the overcoming of social inequalities.

\bigskip





\bibliographystyle{model3-num-names}
\bibliography{<your-bib-database>}

\begin{thebibliography}{00}



\bibitem{Arthur B. Durlauf S. Lane D.A.} Arthur B., Durlauf S., Lane D.A.: Process and the emergence in the economy. In: Arthur B., Durlauf S., Lane D.A. (Eds.) 
The Economy as an Evolving Complex System II, (Introduction). Westview Press (1997)
\bibitem{Bertotti M.L.} Bertotti, M.L.: Modelling taxation and redistribution: a discrete active  particle kinetic approach. Appl. Math. Comput. \textbf{217}, 752--762 (2010)
\bibitem{Bertotti M.L. Modanese G. 1} Bertotti, M.L., Modanese G.: From microscopic taxation and redistribution models to macroscopic income distributions. 
Physica A \textbf{390}, 3782--3793 (2011)
\bibitem{Bertotti M.L. Modanese G. 2} Bertotti, M.L., Modanese G.: Exploiting the flexibility of a family of models for taxation and redistribution. 
Eur. Phys. J.B \textbf{85}, 261 (2012)
\bibitem{Bertotti M.L. Modanese G. 3} Bertotti, M.L., Modanese G.: Mathematical models for socio-economic problems. 
In: Celletti A., Locatelli U., Ruggeri T., Strickland E. (Eds.) 
Mathematical Models and Methods for Planet Earth. Springer INDAM Series 6,
in print,
(2014)
\bibitem{Bloomquist K.M.} Bloomquist K.M.: A Comparison of Agent-Based Models of Income Tax Evasion. Social Science Computer Review  \textbf{24}, 411--425 (2006)
\bibitem{Chakraborti A. Chakrabarti B.K.} Chakraborti A., Chakrabarti B.K.: Statistical mechanics of money: 
how saving propensity affects his distribution. Eur. Phys. J. B, \textbf{17}, 167--170 (2000)
\bibitem{Chatterjee A. Yarlagadda S. Chakrabarti B.K.} Chatterjee A., Yarlagadda S., Chakrabarti B.K. (Eds.): Econophysics of Wealth Distributions. Springer (2005)
\bibitem{Dragulescu A. Yakovenko V.M.} Dragulescu A., Yakovenko V.M.: Statistical mechanics of money. Eur. Phys. J. B, \textbf{17}, 723--729 (2000)
\bibitem{Gallegati M.} Gallegati M.: Reconstructing economics: Agent Based Models and Complexity. 
Conference Paper, Berlin, Paradigm Lost: Rethinking economics and Politics (2012).
Available at {http://ineteconomics.org/conference/berlin/reconstructing-economics-agent-based-models-and-complexity}
\bibitem{Hokamp S. Pickhardt M.} Hokamp S. Pickhardt M.: Income tax evasion in a society of heterogeneous agents: evidence from an agent-based model. International
Economic Journal \textbf{24}, 541--553 (2010)
\bibitem{Kaniadakis G.} Kaniadakis G.: Non-linear kinetics underlying generalized statistics.
Physica A, \textbf{296}, 405--425 (2001)
\bibitem{Kirman A.} Kirman A.: Complex Economics: Individual and Collective Rationality. Routledge, London (2010) 
\bibitem{Mantegna R.N. Stanley H.E.} Mantegna R.N., Stanley H.E.: An Introduction to Econophysics. Cambridge University Press, Cambridge (2000)
\bibitem{Mittone L.} Mittone L.: Dynamic behaviour in tax evasion: an experimental approach. The Journal of Socio-Economics \textbf{35}, 813--835 (2006)
\bibitem{Pareto V.} Pareto V.: Course d'economie politique, Lausanne (1896-7).   
\bibitem{Schelling T.C.} Schelling T.C.: Micromotives and Macrobehavior. Revised edition (First edition in 1978). W.W. Norton e Company, New York (2006)
\bibitem{Sinha S. Chakrabarti B.K.} Sinha S., Chakrabarti B.K.: Towards a physics of economics. 
Physics News (Bullettin of the Indian Physical Association) {\bf 39}, 33--46 (2009)
\bibitem{Yakovenko V.M.} Yakovenko V.M.: Econophysics: Statistical mechanics approach to. 
In: Meyers R.A. (ed.), Encyclopedia of Complexity and System Science, 2800--2826. Springer (2009)
\bibitem{Zaklan G. Westerhoff F. Stauffer D.} Zaklan G., Westerhoff F., Stauffer D.: Analysing tax evasion dynamics via the Ising model. 
J. Econ. Interact. Coord. \textbf{4}, 1--14 (2009) 



\end{thebibliography}



\end{document}